\documentclass[11pt]{article}
\usepackage{amsfonts}
\usepackage{amssymb}
\usepackage{amsmath}
\usepackage{amsthm}
\newlength{\bredde}
\def\slash#1{\settowidth{\bredde}{$#1$}\ifmmode\,\raisebox{.15ex}{/}
\hspace*{-\bredde} #1\else$\,\raisebox{.15ex}{/}\hspace*{-\bredde} #1$\fi}
\newtheorem{theorem}{Theorem}
\newtheorem{lemma}[theorem]{Lemma}
\newtheorem{proposition}[theorem]{Proposition}

\theoremstyle{definition}

\newtheorem{remark}[theorem]{Remark}

\newcommand{\ud}{\mathrm{d}}

\newcommand{\be}{\begin{equation}}
\newcommand{\ee}{\end{equation}}
\newcommand{\bea}{\begin{eqnarray}}
\newcommand{\eea}{\end{eqnarray}}
\newcommand{\nn}{\nonumber}

\newcommand{\hs}{\hat{\sigma}}

\newcommand{\Ai}{\mbox{Ai}}
\newcommand{\sect}[1]{\setcounter{equation}{0}\section{#1}}

\def\Tr{{\mbox{Tr}}}

\begin{document}
\topmargin -1.4cm
\oddsidemargin -0.8cm
\evensidemargin -0.8cm
\title{\Large{{\bf
Interpolation between Airy and Poisson statistics for unitary chiral non-Hermitian
random matrix ensembles
}}}

\vspace{1.5cm}

\author{~\\{\sc G. Akemann}$^{1,2}$ and {\sc M. Bender}$^{3}$\\~\\
$^1$Department of Mathematical Sciences \& BURSt Research Centre\\
Brunel University West London,  Uxbridge UB8 3PH, United Kingdom\\~\\
$^2$The Niels Bohr Institute \& Niels Bohr International Academy\\
Blegdamsvej 17, 2100 Copenhagen {\O}, Denmark\\~\\
$^3$Department of Mathematics, Katholieke Universiteit Leuven\\
Celestijnenlaan 200B, 3001 Leuven, Belgium
}
\maketitle
\vfill
\begin{abstract}
We consider a family of chiral non-Hermitian Gaussian
random matrices in the unitarily invariant symmetry
class. The eigenvalue distribution
in this model is
expressed in terms of Laguerre polynomials in the complex plane.
These are orthogonal with respect to a non-Gaussian weight
including a modified Bessel function of the second kind, and we give an elementary proof for this.
In the large $n$ limit, the eigenvalue statistics at the spectral edge close to the real axis
are described by the same family of kernels interpolating between Airy and Poisson that was
recently found by one of the
authors for the elliptic Ginibre ensemble.
We conclude that this scaling limit is universal, appearing for two different
non-Hermitian random matrix ensembles with unitary symmetry.
As a second result we give an equivalent form for the interpolating Airy
kernel in terms of a single real integral, similar
to representations for the asymptotic kernel
in the bulk and at the hard edge of the spectrum.
This makes its structure
as a one-parameter deformation of the Airy kernel more transparent.
\end{abstract}
\vfill

\thispagestyle{empty}
\newpage

\renewcommand{\thefootnote}{\arabic{footnote}}
\setcounter{footnote}{0}

\sect{Introduction}\label{intro}

Random Matrix Theory (RMT) is 
a rich field of probability theory, with many
applications in physics, mathematics and beyond. One of the results that
exemplifies this success is the appearance of the so-called Tracy-Widom (TW) distribution, describing the
largest eigenvalue of random matrices,
\cite{TW92}, in a large range of contexts,
including combinatorics, growth processes and random tilings,  as reviewed in
\cite{TW02}.
The TW distribution function, which can be expressed 
in terms of a solution to the Painlev\'e II equation, is defined as a Fredholm determinant of
the Airy kernel. This kernel is universal, in the sense that it
describes the soft edge eigenvalue scaling limit for a large class of
non-Gaussian and non-invariant ensembles (see
\cite{TW02} for references).

In the recent past, non-Hermitian RMT has become an area of intensive research,
with applications ranging from the fractional quantum Hall effect \cite{PdF}
to quantum chromodynamics \cite{Steph}, and we refer to \cite{KS09} for a
recent review and more applications. The following question thus naturally
arises: what happens to the largest eigenvalue of a random matrix when it is
allowed to move out into the complex plane?
This question has been answered by one of the authors in \cite{Bender}
for the so-called elliptic Ginibre ensemble with unitary symmetry
($\beta=2$). This is a
Gaussian matrix model generalizing the original complex Ginibre ensemble \cite{Gin}.
It can most conveniently be defined as a two-matrix
model of a Hermitian matrix $H$ and an anti-Hermitian matrix $A$,
depending on a
non-Hermiticity parameter $\tau\in[0,1)$, see e.g. \cite{FKS}.
Depending on the way $\tau$ is scaled when
the matrix size $n$ tends to infinity, a new interpolating behaviour between the Tracy-Widom ($\tau=1$) and Gumbel distributions was found in
\cite{Bender}.
More generally, the limiting two-dimensional eigenvalue point process at the spectral edge
interpolates between Airy and Poisson, and is described by a new
family of kernels, generalizing the Airy kernel.

In this work we address two questions: how universal is this
interpolating Airy kernel, and can its relation to the real Airy kernel be made more
transparent?

To partly answer the first question we consider the chiral extension of the
elliptic Ginibre ensemble with unitary symmetry, introduced in
\cite{Osborn}. Non-Hermitian models with chiral symmetry can typically be
formulated as two-matrix models with two complex
non-Hermitian matrices
(depending again on a non-Hermiticity parameter). In the Gaussian case studied here, the model can be considered as
a non-Hermitian generalization of the  Wishart-Laguerre ensemble. 
In all three symmetry classes  (real non-symmetric ($\beta=1$),
complex  ($\beta=2$) and quaternion
real  ($\beta=4$)), of  chiral non-Hermitian RMTs,
all eigenvalue correlation functions for  finite $n$ can be
expressed in terms of a kernel of
Laguerre polynomials in the complex plane, see
\cite{APSo,Osborn,A05} respectively.
We will present a proof that, for $\beta=2$ in the appropriate large $n$ limit,
the {\it same} interpolating Airy kernel as in \cite{Bender} is found.
In that sense the interpolating Airy kernel is universal, appearing for two
different symmetry classes of
Gaussian non-Hermitian models with unitary invariance.
This is not unexpected as it parallels the
situation for Hermitian ensembles, where chiral and non-chiral models with the
same $\beta$ share the same soft edge spectral behaviour \cite{PF93}.
Our method is similar to that of \cite{Bender}, and involves
the asymptotic analysis of a double
contour integral representation of the Laguerre polynomials.

To answer the second question
we will show that the interpolating Airy kernel permits a
real integral representation as an alternative to the complex double contour integral
representation given in \cite{Bender}. This explicitly reveals the analogy of its structure
as a one-parameter deformation of the Airy kernel with the known non-Hermitian generalizations of the sine and Bessel kernels.

In order to put our results in a more general context, let us briefly review
the different scaling regimes in non-Hermitian RMT, and what is known about
their universality.
The first region that was studied was
the so-called bulk of the support of the limiting eigenvalue
distribution. Here different behaviours occur depending on how the non-Hermiticity parameter scales with $n$; in the "weakly non-Hermitian" regime, $n(1-\tau)=:\hat{\sigma}_n^2\to \hat{\sigma}^2 \in (0,\infty)$,
a transition takes place between "strongly non-Hermitian" statistics ($\hat{\sigma}_n^2 \to \infty$) and what we will call "essentially Hermitian" statistics ($\hat{\sigma}_n^2 \to 0$), described by the sine kernel, see \cite{FKS, FKS98}. 
 Typically, in the strong non-Hermiticity
limit, the correlations can be
obtained from maximal non-Hermiticity ($\tau=0$) by a simple rescaling of the
eigenvalues, see e.g. \cite{FKS98}.
In the strong non-Hermiticity
limit, universality in the bulk was proved rigorously for
general normal random matrices in the $\beta=2$  case in \cite{berman}.
It remains an open problem how to amend the Riemann-Hilbert approach,
which has proven so successful in Hermitian RMT, to
models with complex eigenvalues, see \cite{IT} for a discussion on this.
At weak non-Hermiticity a heuristic argument in favour of universality
was given in \cite{FKS98} and \cite{Ake02} for iid matrix elements and
weight functions with non-Gaussian harmonic potentials  respectively.

Chiral models show a different (hard edge) spectral behaviour at the origin. It depends on the finite multiplicity $\nu$ of the eigenvalue at the origin, which is kept fixed when $n$ goes to infinity.
Again the same transitional regime of weak non-Hermiticity occurs, as identified in \cite{A03}, following \cite{FKS}. At the
origin in this regime, universality is known to hold in the sense that
the same Bessel density in the complex plane is obtained from two different
Gaussian one- and two matrix models, see \cite{SV03} and \cite{Osborn},
respectively.

At the edge of the spectrum close to the real axis
which is our concern here, the appropriate
weakly non-Hermitian scaling limit was identified in \cite{Bender} (see also \cite{GNV}).
Here the transition occurs in a different regime, $\sigma_n^2=(1-\tau_n)n^{1/3}\to (0,\infty)$, when the bulk and hard edge statistics are still strongly non-Hermitian. 
Our work is the first argument presented to suggest the
universality of the corresponding interpolating Airy kernel, as we
obtain it from a second Gaussian model in a different symmetry class.

Let us also mention the decay of the eigenvalue density at the spectral edge in the limit
of strong non-Hermiticity. For all three symmetry classes of the
elliptic Ginibre ensembles ($\beta=1,2,4$) the
density decays as a complementary error function, as was shown in
\cite{KS09} (and references therein). In that sense this behaviour is
also universal. Note that this is consistent with the Poisson statistics of the extreme eigenvalues, which is found on the microscopic scale and shifted out from the spectral edge, as opposed to the mesoscopic scale where the decay of the eigenvalue density appears. 

Our paper is organized as follows. In Section \ref{results},
we briefly introduce the model and summarize our
results, which are: An elementary proof for the orthogonality of the Laguerre
polynomials on $\mathbb C$ which is detailed in Section \ref{OP} (for
the corresponding result for Hermite polynomials, see \cite{PdF,KS09}).
An alternative representation of the kernel of Laguerre polynomials as a
double contour integral shown in Section \ref{kerrep}, which
serves as a Lemma to
prove the convergence in the scaling limit of the eigenvalue process to the interpolating Airy process in Section
\ref{soft}. Finally, we provide a simple real integral representation of the interpolating Airy kernel
  as an alternative to the previously known
double contour integral form; the proof is given in Section \ref{proofla3}. Here we also check that the complementary error function density decay is obtained from the interpolating Airy kernel in the appropriate strongly non-Hermitian scaling limit.


\sect{Summary of results
}\label{results}

We consider a chiral two-matrix model in the unitary
symmetry class which
is the non-Hermitian extension of the chiral Gaussian unitary
ensemble.
For any  non-negative integer $\nu$, let $P$ and $Q$ be $n\times (n+\nu)$ matrices with iid centered complex Gaussian entries of variance 1/(4n),
\be \nn
\ud \tilde{\mathbb{P}}_{n,\nu}(P,Q)=
C_{\nu,n} \ \exp \left[-2n\Tr(PP^\dag+QQ^\dag)\right]\
\ud P
\ud Q
 ,
\ee
 where $\ud P$ denotes Lebesgue measure on the space of complex $n\times (n+\nu)$ matrices (identified with $\mathbb{R}^{2n(n+\nu)}$).
We are interested in the eigenvalue distribution of the $(2n+\nu)\times (2n+\nu)$ random
Dirac matrix
\be
{\cal{D}}=\left(\begin{array}{cc}
0& \Phi\\
 \Psi &0\\
\end{array}
\right)
:= \left(\begin{array}{cc}
0& \sqrt{1+\tau} \ P+\sqrt{1-\tau} \  Q\\
 \sqrt{1+\tau}\ P^\dag -\sqrt{1-\tau}\  Q^\dag&0\\
\end{array}
\right)
\ ,
\label{Diracdef}
\ee
where $\tau\in [0,1]$ is a non-Hermiticity parameter.
Equivalently, 
for $\tau \ne 1$, the distribution of ${\cal{D}}$ is given by the probability measure
\begin{equation}\label{ZPQ}
\ud \tilde{\mathbb{P}}_n^{\tau,\nu}({\cal{D}})
=\frac{1}{\tilde{{\cal Z}}_{n}^{\tau,\nu}}\ \exp\left[-\frac{n}{(1-\tau)}\Tr\left( \Phi\Phi^\dag+ \Psi^\dag \Psi-\tau \left(\Phi \Psi+\Psi^\dag\Phi^\dag \right)\right)\right]\
\ud \Phi
\ud \Psi,
\end{equation}
with $\ud \Phi$ and $\ud \Psi$ denoting Lebesgue measure on the spaces of complex matrices of appropriate dimensions. 
The spectrum of ${\cal{D}}$ consists of an eigenvalue of multiplicity $\nu$ at the origin,  and $2n$ complex eigenvalues $\left\{\pm
 z_k\right\}_{k=1}^{n}$ which come in pairs with opposite sign. Without loss of generality, we choose the $z_k$ to have non-negative real part.  With probability $1$, the $z_k$ are all distinct and have strictly positive real part, and we consider this situation from now on.
Changing variables and integrating out the eigenvectors induces a probability distribution on $\left\{
 z_k\right\}_{k=1}^{n}$,
\be
\ud \mathbb{P}_n^{\tau,\nu}(z_1,\ldots, z_n)=\frac{1}{{\cal Z}_n^{\tau,\nu}}\ |\Delta_n(z^2)|^2\prod_{k=1}^n \,w_{n}^{\tau,\nu}(z_k^2)\chi_{\{\operatorname{Re}z_k > 0\}}\ud^2z_k
,
\label{Zev}
\ee
where $\ud^2 z=\ud x \ud y$ denotes Lebesgue measure in the plane and ${\cal Z}_n^{\tau, \nu}$ is a normalizing constant. Here
we have introduced the Vandermonde determinant
\be \nn
\Delta_n(z):\ = \prod_{1\leq j<k\leq n}(z_k-z_j)\ ,
\ee
and the real and positive weight function
\be
w_{n}^{\tau,\nu}(z):\ = |z|^{\nu+1}
\exp\left[\frac{2\tau n \operatorname {Re}(z)}{(1-\tau^2)}
\right]
K_\nu\left( \frac{2n|z|}{(1-\tau^2)}\right) 
\label{weight}
\ee
where $K_\nu$ is the modified Bessel function of the second kind (also called the MacDonald function).
The limiting case $\tau=1$ reduces the
two-matrix model to the standard Hermitian chiral Gaussian unitary ensemble, which is equivalent to the unitary Wishart-Laguerre ensemble of non-negative definite Hermitian matrices $P P^\dag$ with non-zero eigenvalues $\{z_j^2\}_{j=1}^n$.
For details on the model, we refer to \cite{Osborn} with  $\mu=\sqrt{(1-\tau)}/\sqrt{(1+\tau)}$,  ${\cal D}_{II}=i {\cal{D}} $,  $\alpha=2/(1+\tau)$, $A=\sqrt{1+\tau}P$, $B=i\sqrt{1+\tau}Q$, $C=i\Phi$ and $D=i\Psi$.


The eigenvalue measure (\ref{Zev}) has the structure of a determinantal point process   and
as was shown in \cite{Osborn},
the model can be solved
by introducing Laguerre polynomials orthogonal
in the complex
plane with respect to the weight function (\ref{weight}). The
  orthogonality of the Laguerre polynomials on $\mathbb{C}$
was conjectured in \cite{A03}, however, the proposed weight
  is only correct for $\nu=1/2$.
Let $L_j^{\nu}$ denote
the ordinary generalized Laguerre polynomial with parameter $\nu$ of degree $j$, defined by the orthogonality relation
\begin{equation*}
\int_{0}^{\infty} L_j^{\nu}(x)L_k^{\nu}(x)
x^{\nu}e^ {-x} \ud x
=\frac{(j+\nu)!}{j!}\delta_{jk}.
\end{equation*}

\begin{proposition}\label{laguerre}

For any integers $j,k,\nu \geq 0$  and real numbers
$a>b>0$ the following orthogonality relation holds.
\be\label{laguerreOP}
\langle L_j^{\nu},L_k^{\nu}\rangle :\ = \int_{\mathbb{C}}L_j^{\nu}\left(cz\right)
L_k^{\nu}\left(c\overline{z}\right)
|z|^{\nu} e^{b\operatorname{Re}(z)}
K_\nu\left( a|z|\right)\ud^2z
=h_j^\nu\delta_{jk}\ ,
\ee
with squared norm
\be\label{norm}
h_j^\nu=\frac{\pi(j+\nu)!}{a\ j!}\left( \frac{a}{b}\right)^{2j}
\left(\frac{2a}{a^2-b^2}\right)^{\nu+1}
\ee
and $c: \ = (a^2-b^2)/(2b)$.
\end{proposition}

In stating Proposition \ref{laguerre} we have changed back to
unsquared variables over the full complex plane and general real parameters
$a>b>0$. The case of interest to us,  (\ref{weight}),  follows by setting $a=2n/(1-\tau^2)$,
$b=2\tau n/(1-\tau^2)$ leading to $c=n/\tau$.
While (\ref{laguerreOP}) and  (\ref{norm}) were given in \cite{Osborn}
as ``easy (to) verify'', no formal proof was offered.
A first proof of  (\ref{laguerreOP}) was given in \cite{A05}, Appendix A,
by explicit computation of the moments
$\int \ud^2z \ w_{n}^{\tau,\nu}(z^2)L_j^{\nu}\left(cz^2\right)\overline{z}^{2k}$
in terms of hypergeometric functions. A more elementary proof of Proposition
\ref{laguerre} is presented in
Section \ref{OP}.

An alternative proof could be given exploiting the fact that in this
symmetry class the orthogonal polynomials $P_n(z)$ are directly given by the
expectation value of a characteristic polynomial
$P_n(z)=\mathbb{E}[\det({\cal{D}}-zI)]$, see \cite{AV03}.
The latter can then be computed explicitly for the
Gaussian two-matrix model  (\ref{ZPQ}) using Grassmannians, along the lines
of \cite{APSo}.

In the limit $b\to0$
$(\tau\to0)$ the weight $w(z)$ in  (\ref{weight})
becomes rotationally invariant and the proof of Proposition \ref{laguerre} is
straightforward, replacing the Laguerre polynomials by their leading monic
power (and multiplying  (\ref{laguerreOP}) with the appropriate power
$b^{2j}$).

By a standard argument, performing suitable row and column operations on the Vandermonde determinant,  
all $k$-point correlation functions $R_k$
of the point process
 (\ref{Zev})
can be expressed as determinants of a correlation kernel defined in terms of orthogonal polynomials with respect to $w_{n}^{\tau,\nu}(z)/|z|$,
\begin{align*}
R_k(\zeta_1,\ldots,\zeta_k):\ =& \frac{1}{{\cal Z}_n}\frac{n!}{(n-k)!}
 \int
\ud^2\zeta_{k+1}\cdots \int \ud^2 \zeta_{n} \ |\Delta_n(\zeta^2)|^2 \prod_{j=1}^n
w_{n}^{\tau,\nu}(\zeta_j^2)\chi_{\{\operatorname{Re}\zeta_j > 0\}}\nn\\
=& \det
\left[ \left({\cal K}_n^{\nu,\tau}(\zeta_i,\zeta_j)\right)_{1\leq i,j \leq k }\right],
\end{align*}
where, by  (\ref{laguerreOP}),
\be
{\cal K}_n^{\nu,\tau}(\zeta_1,\zeta_2)=\frac{8n^{2+\nu}}{\pi(1-\tau^2)}
\left(w_{n}^{\tau,\nu}(\zeta_1^2)w_{n}^{\tau,\nu}(\zeta_2^2)\right)^{1/2}
\sum_{k=0}^{n-1}\frac{\tau^{2k}k!}{(k+\nu)!}
L_k^{\nu}\left(\frac{n\zeta_1^2}{\tau}\right)
L_k^{\nu}\left(\frac{n\overline{\zeta}_2^2}{\tau}\right)
\label{kerdef}
\ee
if $\operatorname{Re}\zeta_j > 0$, $j=1,2$, and ${\cal K}_n^{\nu,\tau}(\zeta_1,\zeta_2)=0$ otherwise. When taking the large $n$ scaling limit at the edge of the spectrum
the following representation will prove very useful.

\begin{lemma}\label{kernel}
For any integers $n\geq 1$, $\alpha \geq 0$, parameter $\tau\in(0,1)$
and complex numbers
  $z_1\ne 0$ and $z_2$ the identity
\begin{equation}\label{laguerreformula}
\sum_{k=0}^{n-1}\frac{\tau^{2k}k!}{(k+\alpha)!}
L_k^{\alpha}\left(z_1\right)L_k^{\alpha}\left(z_2\right)\
=\frac{\tau^{2n}e^{z_2}}{4\pi ^2 z_1^{\alpha}}
\oint_{\gamma_1}\ud {u}\oint_{\gamma_2}\ud {v}
\left(\frac{{v}({u}-1)}{({v}-1){u}}\right)^{\alpha}\left(\frac{{v}}{{u}}\right)^n
\frac{e^{z_1
\frac{{u}}{({u}-1)}-z_2\frac{{v}}{({v}-1)}}}{(\tau^2{v}-{u})({v}-1)({u}-1)}
\end{equation}
holds. Here $\gamma_1$ is a simple closed contour encircling  the
point ${u}=0$ but not ${u}=1$, while  $\gamma_2$ encircles both the point ${v}=1$
and the entire contour $\gamma_1$ in such a way that $\tau^2{v}-{u}\neq 0$
for ${u}\in \gamma_1$ and ${v}\in \gamma_2$.
\end{lemma}

Now we state our main result, giving  scaling limits for the eigenvalue distribution of ${\cal{D}}$. 
We emphasize that the following result holds for any fixed $\nu$. In principle, our method could be extended to cover the case when $\nu$ scales with $n$ (see Remark \ref{scalealpha}), the difficulty here is to identify the global spectral distribution, a non-Hermitian generalization of the Marcenko-Pastur law.

\begin{theorem}\label{main2} Let $\nu$ be a fixed non-negative integer.
Given a sequence  $\{\tau_n\}_{n=1}^{\infty}\subset [0,1)$, put
$\sigma_{n}=(2n)^{1/6}\sqrt{(1-\tau_n)}$. Let $z_j=x_j+iy_j$, $j=1,\ldots, n$ be the eigenvalues with positive real part of the random Dirac matrix ${\cal{D}}$, forming a determinantal process with kernel ${\cal K}_n^{\nu,\tau_n}$. For the choices
of scaling parameters ${a}_n$, ${b}_n$ and
${c}_n$ specified below, put $\hat x_j =(x_j-c_n)/{a}_n$ and $\hat y_j= y_j/{b}_n$, define the rescaled two-dimensional eigenvalue point process
${Z}_n^{\tau_n, \nu}=
\left\{(\hat x_j,\hat y_j)\right\}_{j=1}^n$,
and let $F_{n}^{\tau_n,\nu}(t)
= \mathbb{P}_{n}^{\tau_n,\nu}\left[\max_{1\leq j\leq n}\{\hat x_j \}\leq t\right]$
be the last particle distribution of ${Z}_n^{\tau_n,\nu}$.
\begin{itemize}
\item[(i)]\label{main2b} Suppose
$\sigma_n\to \sigma \in [0,\infty)$ as $n\to \infty$.
Choose  
${a}_n=(2n)^{-2/3}$, ${b}_n=\sigma_n (2n)^{-2/3}$
and
$c_n=(1+\tau_n)$.
Then ${Z}_n^{\tau_n,\nu}$ converges weakly to
$Z^A_{\sigma}$,
a determinantal point process on $\mathbb{C} \cong \mathbb{R}^2$ with correlation kernel
\be
\label{Airyrep1}
{\cal K}_{\sigma}^A(\zeta_1,\zeta_2) :\ =
\frac{1}{4\pi^{5/2}}\int_{\mathbb{R}+i\delta}\ud u\int_{\mathbb{R}+i\delta}\ud v
\frac{e^{-\frac{1}{2}(\sigma
    u+\eta_1)^2+\frac{i}{3}u^3+i\xi_1u-\frac{1}{2}(\sigma
    v-\eta_2)^2+\frac{i}{3}v^3+i\xi_2v}
}{i(u+v)},
\ee
where $\zeta_j=\xi_j+i \eta_j$,
and $F_{n}^{\tau_n,\nu}(t)$ converges to the last particle distribution
$F_{\sigma}(t)$ of $Z^A_{\sigma}$. 

\item[(ii)]\label{main1} Suppose $\sigma_n \to \infty$ as $n\to \infty$.
Put $\hat{\tau}_n=(1+\tau_n)/2$ and choose
\begin{equation}{a}_n\label{gumbelreal}
=\sqrt{\hat{\tau}_n}\frac{\sigma_n }{\sqrt{6 \log \sigma_n}} (2n)^{-2/3},
\end{equation}

\begin{equation}
\label{gumbelimaginary}{b}_n
={\hat \tau_n}^{-1/4}\frac{\sigma_n^{5/2}}{(6 \log \sigma_n)^{1/4}}(2n)^{-2/3},
\end{equation}
and
\begin{equation}
\label{gumbelmean}c_n
=(1+\tau_n)+{a}_n
\left(3\log \sigma_n-\frac{5}{4}\log( 6\log \sigma_n)
-\log(2\pi{\hat\tau}^{3/4})\right).
\end{equation}
Then
${Z}_n^{\tau_n,\nu}$
converges weakly to a Poisson process
on $\mathbb{R}^2$ with intensity $\pi^ {-1/2}e^{-\xi-\eta^2}$,
and
$F_{n}^{\tau_n, \nu}(t)\to e^{-e^{-t}}$,  the Gumbel distribution.
\end{itemize}
\end{theorem}
The interpolating Airy kernel (\ref{Airyrep1})  was derived as the
scaling limit for the corresponding non-chiral Gaussian ensemble
in \cite{Bender}. We show that this kernel has an alternative real integral
representation.  To clarify the analogy with the transitional Bessel and sine kernels below, we consider the scaling limit corresponding to $b_n=a_n=(2n)^{-2/3}$ rather than $b_n=\sigma_n a_n=\sigma_n (2n)^{-2/3} $ in the first part of Theorem \ref{main2};
for $\sigma >0$, define the rescaled version
  \begin{align}\label{Airyrep1b}
  \hat {\cal K}_{\sigma}^A(\zeta_1,\zeta_2):=&\frac{1}{\sigma} {\cal K}_{\sigma}^A(\xi_1+i\eta_1/\sigma,\xi_2+i \eta_2/\sigma)\nn\\
  =&
  \frac{e^{-\frac{1}{2\sigma^2}(\eta_1^2+\eta_2^2)}}{4\pi^{5/2}\sigma}\int_{\mathbb{R}+i\delta}\ud u\int_{\mathbb{R}+i\delta}\ud v \
\frac{e^{-\frac{1}{2}\sigma^2
    u^2+\frac{i}{3}u^3+i\zeta_1 u-\frac{1}{2}\sigma^2
    v^2+\frac{i}{3}v^3+i\overline{\zeta}_2 v}
}{i(u+v)}
\end{align}
of the kernel (\ref{Airyrep1}).
\begin{proposition}\label{Airyint}
 The interpolating Airy kernel has the real integral representation
\begin{equation}\label{Airyrep2}
\hat {\cal K}_{\sigma}^A(\zeta_1,\zeta_2)
=\frac{e^{-\frac{1}{2\sigma^2}(\eta_1^2+\eta_2^2)+\frac16\sigma^6
+\frac{\sigma^2}{2}\left(\zeta_1 +\overline{\zeta}_2\right)}}{\sigma\sqrt{\pi}}
\int_{0}^\infty \ud t\ e^{t\sigma^2}\
{\rm Ai}\left(\zeta_1+\sigma^4/4+t\right)
{\rm Ai}\left(\overline{\zeta}_2+\sigma^4/4+t\right),
\end{equation}
where $\zeta_j=\xi_j+i\eta_j$ and ${\rm Ai}$ is the Airy function.
\end{proposition}
The convergence of the integral in (\ref{Airyrep2})
can easily be seen from the large argument asymptotics of the Airy function,
\begin{equation}\label{Airyasymptotics}
\Ai(z)=\frac{1}{2\sqrt{\pi}}\frac{e^{-\frac23 z^{3/2}}}{z^{1/4}}(1+o(1)),
\end{equation}
as $z\to \infty$ with $\operatorname{Arg}(z)$ bounded away from $\pi$.

In the limit $\sigma\to0$ the integral in
 (\ref{Airyrep2}) becomes elementary (see  (2.37) in \cite{Kurt}
for this integral form),
\be \nn
{\cal K}_{\sigma}^A(\zeta_1,\zeta_2)
\ \stackrel{{\sigma}\to0}{\longrightarrow}\
\frac{e^{-\frac{1}{2}(\eta_1^2+\eta_2^2)}}{\sqrt{\pi}}
\int_{0}^\infty \ud t \
{\rm Ai}\left(\xi_1+t\right){\rm Ai}
\left({\xi}_2+t\right)\ =
\frac{e^{-\frac{1}{2}(\eta_1^2+\eta_2^2)}}{\sqrt{\pi}}
{\cal K}^{A}(\xi_1,\xi_2),
\ee
where ${\cal K}^{A}$
 is the Airy kernel.
Proposition \ref{Airyint} completes the picture of the known non-Hermitian
generalizations
\begin{equation}
{\cal K}_{\hat{\sigma}}^{S}(\zeta_1,\zeta_2)
=\frac{e^{-\frac{1}{2\hs^2}(\eta_1^2+\eta_2^2)}}{\hat{\sigma}\pi^{3/2}}
\int_{0}^1 \ud t \ e^{-t^2\hat{\sigma}^2}
\cos(t(\zeta_1-\zeta_2)),
\end{equation}
\cite{FKS98}
of the sine kernel ${\cal K}^{S}$ and
\begin{equation}
{\cal K}_{\hat{\sigma}}^{B,\nu}(\zeta_1,\zeta_2)
=\frac{|\zeta_1\zeta_2|^{\nu+1}}{2\pi\hs^2(\zeta_1\overline{\zeta_2})^\nu}
\sqrt{K_\nu\left(\frac{|\zeta_1|^2}{4\hs^2}\right)
K_\nu\left(\frac{|\zeta_2|^2}{4\hs^2}\right)}\
e^{\frac{1}{8\hs^2}(\xi_1^2+\xi_2^2)}
\int_{0}^1 \ud t\,t
e^{-2 t^2\hat{\sigma}^2}
J_\nu(t\zeta_1)J_\nu(t\zeta_2),
\end{equation}
 \cite{A03,Osborn}
of the Bessel kernel ${\cal K}^{B,\nu}$, in the bulk and at the hard edge of the
spectrum, respectively.
These are obtained in the limit when  $\hat{\sigma}= \lim_{n\to\infty}\sqrt{(1-\tau_n)n}  \in (0,\infty)$,
and are both given as simple one-parameter deformations of the ordinary sine and Bessel kernels, which are recovered in the essentially Hermitian limit,
\bea
\pi\hat{\sigma}{\cal K}_{\hat{\sigma}}^{S}(\pi\xi_1+i\hat{\sigma}\eta_1,\pi\xi_2+i\hat{\sigma}\eta_2)
&\stackrel{\hat{\sigma}\to0}{\longrightarrow}&\
\frac{e^{-\frac{1}{2}(\eta_1^2+\eta_2^2)}}{\sqrt{\pi}}
\frac{\sin(\pi(\xi_1-\xi_2))}{\pi(\xi_1-\xi_2)}
=\frac{e^{-\frac{1}{2}(\eta_1^2+\eta_2^2)}}{\sqrt{\pi}}{\cal K}^{S}(\xi_1,\xi_2), \nn \\
\sqrt{2}\hat{\sigma}{\cal K}_{\hat{\sigma}}^{B,\nu}(\xi_1+i\sqrt{2}\hat{\sigma}\eta_1,\xi_2+i\sqrt{2}\hat{\sigma}\eta_2)
&\stackrel{\hat{\sigma}\to0}{\longrightarrow}&
\frac{e^{-\frac{1}{2}(\eta_1^2+\eta_2^2)}}{\sqrt{\pi}}\sqrt{|\xi_1\xi_2|}
\frac{J_\nu(\xi_1)\xi_2J_{\nu+1}(\xi_2)-
J_\nu(\xi_2)\xi_1J_{\nu+1}(\xi_1)}{\xi_1^2-\xi_2^2}\nn\\
&=&\frac{e^{-\frac{1}{2}(\eta_1^2+\eta_2^2)}}{\sqrt{\pi}} {\cal K}^{B,\nu}(\xi_1,\xi_2), \nn \ \ \ \ \
\eea
where we have rescaled the imaginary part with a factor $\hat \sigma$ to get a finite limit, analogously with the definition (\ref{Airyrep1}) of ${\cal K}_{\sigma}^A$.
Note that the bulk and hard edge transitions occur in a different regime $(1-\tau_n) \asymp n^{-1}$ compared to the soft edge $(1-\tau_n) \asymp n^{-1/3}$.


\begin{remark}
From the representation (\ref{Airyrep2}) it is also transparent that our results are consistent with the expected error function decay of the eigenvalue density at the spectral edge when $n \to \infty$; the density of rescaled  eigenvalues $\hat z _j:=\sqrt{2n}(1-\tau_n)^{-1/2}(z_j-(1+\tau_n))$ is given by
\begin{equation}\label{errorfunctiondecay}
\rho_n(\zeta):\ =\frac{(1-\tau_n)}{2n}{\cal K}_n^{\nu,\tau_n}\left( (1+\tau_n)+\sqrt{\frac{(1-\tau_n)}{2n}}\zeta,(1+\tau_n)+\sqrt{\frac{(1-\tau_n)}{2n}}\zeta\right)\to \frac{1}{2\pi(1+\tau)} \operatorname{erfc}(\xi),
\end{equation}
where $\zeta=\xi+i \eta$, $\tau=\lim_{n \to \infty}\tau_n$, and
$\operatorname{erfc}(u)=\frac{2}{\sqrt{\pi}}\int_{u}^\infty \ud t \ e^{-t^2}$ is the complementary error function. See Section \ref{proofla3} for a sketch of this argument.
\end{remark}

\sect{Proof of Proposition \ref{laguerre}}\label{OP}
In this section we present an elementary proof of the orthogonality of the Laguerre polynomials
in the complex plane, Proposition
\ref{laguerre}, using induction in $\nu$. We will make use of known recursions for the Laguerre polynomials and modified Bessel functions which are of depth 2, so we first show the base cases $\nu=0$ and $\nu=1$.

To that end we will use an integral representation for the modified Bessel
function of the second kind, (\cite{Grad},  8.432.6),
\be
K_\nu(x)=\frac12\left(\frac{x}{2}\right)^\nu \int_0^\infty \frac{\ud
  t}{t^{\nu+1}} e^{-t-x^2/(4t)} \ ,
\label{Kint}
\ee
as well as a complex contour integral representation for the generalized Laguerre
polynomials
\be
L_j^\nu(x)=\frac{1}{2\pi i}\oint_{\gamma}\ud u
\frac{e^{-xu/(1-u)}}{(1-u)^{\nu+1}u^{j+1}}\,  
\label{Lint1}
\ee
where the contour $\gamma$ encircles the point $u=0$ but not $u=1$.
We will choose $\gamma$ to be a circle of radius $r<b/a$, centered at the origin.
For $\nu=0$,  (\ref{Lint1}) is given for instance in
\cite{Weber}, equation 13.53. Alternatively this relation can easily be
  derived using the generating function for Laguerre polynomials.
 The case $\nu=1$ (and indeed any positive integer $\nu$, although not needed here) then follows easily using
the identity $L_{n}^{\nu+1}(x)=
-L_{n+1}^{\nu \ \ '}(x)$.

Inserting the representations (\ref{Kint}) and (\ref{Lint1}) with $\nu=0$
into (\ref{laguerreOP}) gives
\begin{align}
\langle L_j^{0},L_k^{0}\rangle=
& \int_{-\infty}^{\infty}\ud x\int_{-\infty}^{\infty}\ud y\ e^{bx}
\int_0^\infty \frac{\ud
  t}{2 t} e^{-t-\frac{a^2}{4t}(x^2+y^2)}
\frac{1}{(2\pi i)^2}\oint_{\gamma}\ud u\frac{e^{-c(x+iy)u/(1-u)}}{(1-u)u^{j+1}}
\oint_{\gamma}\ud v\frac{e^{-c(x-iy)v/(1-v)}}{(1-v)v^{k+1}}
\nn\\
=& \frac{1}{2(2\pi i)^2}\int_{-\infty}^{\infty}\ud x\int_{-\infty}^{\infty}\ud y
\int_0^\infty \ud t\oint_{\gamma}\ud u
\oint_{\gamma}\ud v \ \frac{e^{-tC
-\frac{a^2}{4t}(x+tA)^2-\frac{a^2}{4t}(y+tB)^2} }{t(1-u)u^{j+1}(1-v)v^{k+1}},
\label{OPstep1}
\end{align}
where
\begin{equation*}
A=\frac{2}{a^2}
\left(c\left(\frac{u}{1-u}+\frac{v}{1-v}\right)-b\right),
\end{equation*}
\begin{equation*}
B=\frac{2ic}{a^2}
\left(\frac{u}{1-u}-\frac{v}{1-v}\right),
\end{equation*}
and
\begin{equation*}
C=1-\frac{a^2}{4}(A^2+B^2)
=
\frac{(a^2-b^2)}{a^2(1-u)(1-v)}\left(1-\frac{a^2}{b^2}uv \right)
\end{equation*}
depend only on the variables $u$ and $v$.
Note that $\operatorname{Re}(C)>0$ for any  $u,v$ on $\gamma$ since $|u|=|v|=r<b/a<1$, by our choice of $\gamma$.
This means that we can apply Fubini's theorem to calculate first the Gaussian
integrals in $x$ and $y$ and then the convergent
$t$-integral.
Inserting all this into (\ref{OPstep1}) gives
\begin{align*}
\langle L_j^{0},L_k^{0} \rangle =&
 \frac{2\pi}{(a^2-b^2)}\frac{1}{(2\pi i)^2}\oint_{\gamma}\frac{\ud u}{u^{j+1}}
\oint_{\gamma}\frac{\ud v}{v^{k+1}}
\frac{1}{(1-\frac{a^2}{b^2}uv)}\nn\\
=& \frac{2\pi}{(a^2-b^2)}\left(\frac{a^2}{b^2}\right)^k\frac{1}{2\pi i}
\oint_{\gamma}\frac{\ud u}{u^{j-k+1}}\nn\\
=&\frac{2\pi}{(a^2-b^2)}\left(\frac{a^2}{b^2}\right)^k\delta_{jk}\nn\\
=&
h_j^0 \delta_{jk},
\end{align*}
proving the base case $\nu=0$.

The result for $\nu=1$ follows along the same lines and we only give some
intermediate steps. A straightforward calculation using integration by parts shows that, as long as $\operatorname{Re}(C)>0$,
\begin{equation}
\int_{-\infty}^{\infty}\ud x\int_{-\infty}^{\infty}\ud y
\int_0^{\infty} \ud t \frac{(x^2+y^2)}{t^2}e^{-C t-\frac{a^2}{4t} (x+tA)^2
-\frac{a^2}{4t}(y+t B)^2}=\frac{4\pi}{a^2 C}\left(\frac{4}{a^2}+\frac{A^2+B^2}{C}\right)=\frac{16\pi}{a^4C^2},
\end{equation}
implying that
\begin{align*}
\langle L_j^{1},L_k^{1} \rangle
&=
\frac{a}{4 }\frac{1}{(2\pi i)^2}\int_{-\infty}^{\infty}\ud x\int_{-\infty}^{\infty}\ud y
\int_0^\infty \ud t \oint_{\gamma}\ud u\oint_{\gamma}\ud v \
\frac{(x^2+y^2)e^{-tC-\frac{a^2}{4t}(x+tA)^2
-\frac{a^2}{4t}(y+tB)^2
}}{t^2(1-u)^2u^{j+1}(1-v)^2v^{k+1}}
\nn\\
&=\frac{4a\pi}{(a^2-b^2)^2}
\frac{1}{(2\pi i)^2}\oint_{\gamma}\frac{\ud u}{u^{j+1}}
\oint_{\gamma}\frac{\ud v}{v^{k+1}}
\frac{1}{(1-(a/b)^2uv)^2}\nn\\
&=\frac{4a\pi}{(a^2-b^2)^2}\left(\frac{a^2}{b^2}\right)^j(j+1)\delta_{jk}\nn\\
&=h_j^1\delta_{jk},
\end{align*}
since
\begin{equation*}
\frac{1}{2\pi i}\oint_{\gamma}\frac{\ud v}{v^{k+1}(1-(a/b)^2uv)^2}
=
\operatorname{Res}_{v=0}\left(v^{-k-1}\left(\sum_{j=0}^{\infty}\left(\frac{a^2u}{b^2}\right)^jv^j\right)^2\right)
=\sum_{j=0}^k\left(\frac{a^2u}{b^2}\right)^k=(k+1)\left(\frac{a^2u}{b^2}\right)^k.
\end{equation*}

Next we can make the induction step using the recurrence relation for the
modified Bessel functions of the second kind, (\cite{Grad}, 8.486.10),
\begin{equation}
xK_{\nu+1}(x)=xK_{\nu-1}(x)+2\nu K_\nu(x)\ ,
\label{Krec}
\end{equation}
as well as the following two recurrences for Laguerre polynomials, (\cite{Grad}, 8.974.3, 8.971.2 and 8.971.5 respectively)
\begin{equation}
L_n^{\nu+1}(x)=\sum_{m=0}^nL_m^\nu(x)\ ,
\end{equation}
and
\begin{equation}
xL_n^{\nu+1}(x) = \nu L_n^\nu(x)-(n+1)L_{n+1}^{\nu-1}(x)= \nu \sum_{m=0}^nL_m^{\nu-1}(x)-(n+1)L_{n+1}^{\nu-1}(x)
\ .
\label{laguerrerec}
\end{equation}
Suppose (\ref{laguerreOP}) holds for a fixed $\nu$ and $\nu-1$
and every $j,k$. Without loss of generality we consider $j\geq k$. By equations (\ref{Krec}) through (\ref{laguerrerec}) and the induction hypothesis,
\begin{align}
& \langle L_j^{\nu+1},L_k^{\nu+1}\rangle \nn \\
=& \int_{\mathbb{C}} \ud^2z \ e^{b\operatorname{Re}(z)}
L_j^{\nu+1}\left(cz\right) L_k^{\nu+1}\left(c\overline{z}\right)
\left(\frac1a2\nu|z|^{\nu}K_{\nu}\left( a|z|\right)+|z|^2|z|^{\nu-1}K_{\nu-1}\left( a|z|\right)
\right)
\nn\\
=& \int_{\mathbb{C}}\ud^2z \ e^{b\operatorname{Re}(z)}\ \left[\frac{2\nu}{a}
\left( \sum_{m=0}^jL_m^\nu(cz)\right)\left(
\sum_{n=0}^kL_n^\nu(c\overline{z})\right)|z|^{\nu}K_{\nu}\left( a|z|\right)
\right.
\nn\\
&\left.
+\frac{1}{c^2}
\left(\nu \sum_{m=0}^jL_m^{\nu-1}(cz)-(j+1)L_{j+1}^{\nu-1}(cz)\right)\left(\nu \sum_{n=0}^kL_n^{\nu-1}(c\overline{z})
-(k+1)L_{k+1}^{\nu-1}(c\overline{z})\right)|z|^{\nu-1}K_{\nu-1}\left( a|z|\right)
 \right]
\nn\\
=&
\frac{2\nu}{a} \sum_{m=0}^j \sum_{n=0}^k \langle L_m^{\nu},L_n^{\nu}\rangle
+\frac{\nu^2}{c^2}
\sum_{m=0}^j \sum_{n=0}^k \langle L_m^{\nu-1},L_n^{\nu-1}\rangle
-\frac{\nu(j+1)}{c^2}\sum_{n=0}^k  \langle L_{j+1}^{\nu-1},L_{n}^{\nu-1}\rangle \nn\\
&
-\frac{\nu(k+1)}{c^2}\sum_{m=0}^j  \langle L_m^{\nu-1},L_{k+1}^{\nu-1}\rangle
+
 \frac{(j+1)(k+1)}{c^2}\langle L_{j+1}^{\nu-1},L_{k+1}^{\nu-1}\rangle
\nn\\
=&
\frac{2\nu}{a}\sum_{n=0}^kh_n^{\nu}+ \frac{\nu^2}{c^2}\sum_{n=0}^kh_n^{\nu-1}
-\frac{(k+1)\nu}{c^2} h_{k+1}^{\nu-1} (1-\delta_{jk})+
\frac{(k+1)^2}{c^2}h_{j+1}^{\nu-1}\delta_{jk}
.
\label{stepn+1}
\end{align}
Note that the last two terms contribute only when $j>k$ and $j=k$ respectively.
Inserting the expressions (\ref{norm}) we obtain
for the two sums
\begin{align*}
\frac{2\nu}{a}
\sum_{n=0}^kh_n^{\nu}+\frac{\nu^2}{c^2}\sum_{n=0}^kh_n^{\nu-1}
=&\frac{2\nu}{a}\sum_{n=0}^k\frac{\pi(n+\nu)!}{a\ n!}\frac{a^{2n}}{b^{2n}}
\left(\frac{2a}{a^2-b^2}\right)^{\nu+1}\\
&+\frac{\nu^24b^2}{(a^2-b^2)^2}
\sum_{n=0}^k\frac{\pi(n+\nu-1)!}{a\ n!}\frac{a^{2n}}{b^{2n}}
\left(\frac{2a}{a^2-b^2}\right)^{\nu}
\\
=& \frac{4\pi \nu(2a)^\nu}{a (a^2-b^2)^{\nu+2}}\sum_{n=0}^k
\frac{(n+\nu-1)!}{n!}\frac{a^{2n}}{b^{2n}}((n+\nu)a^2-n b^2)\\
=&\frac{4\pi\nu(2a)^\nu}{(a^2-b^2)^{\nu+2}}
\frac{(k+\nu)!}{ k!}\frac{a^{2k+1}}{b^{2k}}\  \\
=&\frac{(k+1)\nu}{c^2} h_{k+1}^{\nu-1}
\end{align*}
after telescoping the sum.
Thus (\ref{stepn+1}) becomes
\begin{align*}
\langle L_j^{\nu+1},L_k^{\nu+1} \rangle =&\frac{(k+1)\nu}{c^2}
 h_{k+1}^{\nu-1}(1-(1-\delta_{jl}))+
\frac{(k+1)^2}{c^2}h_{k+1}^{\nu-1}\delta_{jk}\nn\\
=&\frac{(k+1)}{c^2}(k+1+\nu)h_{k+1}^{\nu-1}\delta_{jk}
\nn\\
=&\frac{\pi(k+\nu+1)!}{a\ k!}\left( \frac{a}{b}\right)^{2k}
\left(\frac{2a}{a^2-b^2}\right)^{\nu+2}\delta_{jk}\nn\\
=& h_k^{\nu+1}\delta_{jk}
\end{align*}
which concludes the proof.

\sect{Proof of Lemma \ref{kernel}}\label{kerrep}


In this section we prove the alternative complex contour integral
representation for the kernel, Lemma \ref{kernel}. To that end we will derive
two different contour integral representations for the Laguerre
polynomials, and insert them
into  (\ref{laguerreformula}). This will obviously break the
symmetry in the arguments $z_1$ and $z_2$.

The first integral representation follows starting from \cite{Abramowitz},
 22.10.7,
\begin{equation}
L_k^{\alpha}(z)=
\frac{e^z}{2\pi i z^{\alpha}}\oint_{\gamma'}\ud w
\frac{w^{k+\alpha} e^{-w}}{(w-z)^{k+1}}.
\end{equation}
valid for any integers $k,\alpha$ and $z\neq0$ (for simplicity we exclude
non-integer indices $\alpha$ to avoid branch cuts). The contour
$\gamma'$ includes the pole at $w=z$ but not the origin $w=0$.

Now make the change of variables $v=w/(w-z)$, $\ud v =-z/(w-z)^2\ud w$,
giving rise to
\begin{equation}\label{hermiteintegral1}
L_k^{\alpha}(z)=\frac{e^z}{2\pi i}\oint_{\gamma_2}\ud v\frac{v^{k+\alpha}
  e^{-zv/(v-1)}}{(v-1)^{\alpha+1}}\ .
\end{equation}
Together with the sign from the Jacobian and the inversion from the M\"obius
transformation,
the contour of integration is mapped to a
positively oriented contour  encircling the pole at $v=1$ but not the origin
$v=0$. Without loss of generality it can be chosen subject to
the second condition in Lemma \ref{kernel} and thus taken as $\gamma_2$.

To prove the second integral representation,
\begin{equation}\label{hermiteintegral3}
\frac{k!L_k^{\alpha}(z)}{(k+\alpha)!}=\frac{1}{2\pi i(-z)^{\alpha}}
\oint_{\gamma_1}\ud u\frac{(1-u)^{\alpha-1}}{u^{k+1+\alpha}}e^{-zu/(1-u)}\ ,
\end{equation}
with $\gamma_1$ as in the statement of the Lemma, 
we use induction over $\alpha$.
The base $\alpha=0$ is given by  (\ref{Lint1}),
\begin{equation*} 
L_k^0(z)
=\frac{1}{2\pi i}\oint_{\gamma_1}\ud u\frac{e^{-zu/(1-u)}}{(1-u)u^{k+1}} \ .
\end{equation*}
Using the known recurrence relation for Laguerre polynomials (e.g. \cite{Grad},
 8.971.4)
\begin{equation}
zL_k^{\alpha+1}(z)=(k+\alpha+1)L_k^{\alpha}(z)
-(k+1)L_{k+1}^{\alpha}(z),
\end{equation}
we can establish the induction step.
Suppose (\ref{hermiteintegral3}) holds for a fixed $\alpha$ and every
$k$. Then for any $k \geq 0$ and $z\neq0$
\begin{align*}
\frac{k!L_k^{\alpha+1}(z)}{(k+\alpha+1)!}
&=\frac{k!L_k^{\alpha}(z)}{z(k+\alpha)!}
-\frac{(k+1)!L_{k+1}^{\alpha}(z)}{z(k+\alpha+1)!}\nn\\
&=\frac{1}{2\pi i z(-z)^{\alpha}}
\oint_{\gamma_1}\ud u\frac{(1-u)^{\alpha-1}}{u^{k+1+\alpha}}e^{-zu/(1-u)}
\left(1-\frac{1}{u}\right)\nn\\
&=\frac{1}{2\pi i(-z)^{\alpha+1}}
\oint_{\gamma_1}\ud u\frac{(1-u)^{\alpha}}{u^{k+1+\alpha+1}}e^{-zu/(1-u)}\ ,
\end{align*}
where we used the induction hypothesis in the first line.

Combining the two representations (\ref{hermiteintegral1})
and (\ref{hermiteintegral3}) gives 
\begin{align}\label{kernelcombi}
\sum_{k=0}^{n-1}\frac{\tau^{2k}k!}{(k+\alpha)!}
L_k^{\alpha}\left(z_1\right)L_k^{\alpha}\left(z_2\right)
&=\sum_{k=0}^{n-1} \frac{\tau^{2k}e^{z_2}}{(2 \pi i) ^2
  (-z_1)^{\alpha}}\oint_{\gamma_2}\ud v\frac{v^{k+\alpha}
  e^{-z_2v/(v-1)}}{(v-1)^{\alpha+1}}
\oint_{\gamma_1}\ud u\frac{(1-u)^{\alpha-1}}{u^{k+1+\alpha}}e^{-z_1u/(1-u)}
\nn\\
&=\frac{e^{z_2}}{(2\pi i)^2
  z_1^{\alpha}}\oint_{\gamma_1}\ud u\oint_{\gamma_2}\ud v
\left(\frac{v(u-1)}{(v-1)u}\right)^{\alpha}
\sum_{k=0}^{n-1} \frac{1}{u}\left(\frac{\tau^2v}{u}\right)^k
\frac{e^{-z_2\frac{v}{(v-1)}
-z_1\frac{u}{(1-u)}}}{(v-1)(1-u)}.
\end{align}
After computing the geometric sum,
\be\label{geometric}
\sum_{k=0}^{n-1}\frac{1}{u}
\left(\frac{\tau^2v}{u}\right)^k=\frac{1}{(\tau^2v-u)}\left(\frac{\tau^2v}{u}\right)^n-\frac{1}{(\tau^2v-u)} ,
\ee
we only have to show that the second term in (\ref{geometric}) does not contribute to
(\ref{kernelcombi}) in order to arrive at the desired result
(\ref{laguerreformula}).
To see this, we change variables $v=y/{(y-1)}$, 
$\ud v=1/(y-1)^2 \ud y $, thereby mapping $\gamma_2$ to a new contour
$\gamma'_2$ which will not enclose the pole at $y=u/(u-\tau^2)$ since $\gamma_2$
does by definition, and the M\"obius transformation $v\mapsto y$ involves an
inversion.
The second term of the integral in (\ref{kernelcombi}) with respect to $y$ then
becomes
\begin{align*}
\oint_{\gamma_2}\ud v
\left(\frac{v}{(v-1)}\right)^{\alpha}
\frac{e^{-z_2\frac{v}{(v-1)}}}
{(\tau^2v-u)(v-1)}&=
\oint_{\gamma'_2}\ud y
\frac{y^{\alpha}e^{-z_2y}}{(\tau^2y/{(y-1)}-u)(y/{(y-1)}-1)
(1-y)^2}\nn \\
&=\oint_{\gamma'_2}\ud y
  \frac{y^{\alpha}e^{-z_2y}}{(y(\tau^2-u)+u)}=0,
  \end{align*}
by Cauchy's theorem.

\sect{Proof of Theorem \ref{main2}}\label{soft}

To prove our main result, we show that (a kernel equivalent to) the  correlation kernel ${\cal K}_n^{\nu,\tau_n}$ of the eigenvalue point process, appropriately rescaled, converges point-wise to the limiting kernel associated with the sequence $\{\tau_n\}$ in the statement of the theorem. The convergence of processes and last particle distributions then follows from a dominated convergence argument, see \cite{Bender}.
We will assume that $\tau_n$ is bounded away from $0$; the case $\tau_n\to 0$ reduces to the (simpler) radially symmetric case where the kernel can be expressed in terms of monomials rather than Laguerre polynomials; our results remain valid in this case.
The analysis is very similar to what was done in great detail in \cite{Bender} for the case of the elliptic Ginibre ensemble, so for clarity we will skip some technical details and refer to those calculations to give an idea of how the present analysis can easily be made completely rigorous.
Our strategy is to use Lemma \ref{kernel} to represent ${\cal K}_n^{\nu,\tau}$
and find its large $n$ asymptotics by steepest descent analysis of the integrals.


It is known that the mean eigenvalue density of ${\cal{D}}$ is asymptotically constant in the ellipse  $\left \{(\xi,\eta): \xi^2/(1+\tau_n)^2+\eta^2/(1-\tau_n) ^2\leq 1 \right\}$. We are interested in the scaling limit around the spectral edge on the positive real axis, so we define the new scaling parameter $\delta_n:=c_n^2-(1+\tau_n)^2 \ll 1$.
The choice of parameters will be determined by the analysis, however in order
to estimate orders of magnitudes of error terms, we will assume when needed that the parameters scale correctly with $n$, that is
\begin{equation*}
a_n\asymp\frac{\sigma_n }{\sqrt{\log \sigma_n}} n^{-2/3},
\end{equation*}
and
\begin{equation*}
{b}_n
\asymp\frac{\sigma_n^{5/2}}{(\log \sigma_n)^{1/4}}n^{-2/3},
\end{equation*}
and
\begin{equation*}
\delta_n\asymp\sigma_n \sqrt{\log \sigma_n}\ n^{-2/3}
\end{equation*}
if $\sigma_n\to \infty$
and $a_n\asymp n^{-2/3}$, $b_n\asymp \sigma_n n^{-2/3}$, $\delta_n\equiv0$ otherwise.

Using Lemma \ref{kernel} to represent the kernel (\ref{kerdef}) and
the large argument asymptotics of the modified Bessel function of the second kind (\cite{Abramowitz}, 9.7.2),
\be \nn
K_{\nu}(z)=\sqrt{\frac{\pi}{2z}}e^{-z}(1+o(1)),
\ee
the kernel ${\cal M}_n^{\nu,\tau_n}(\zeta_1,\zeta_2):={a}_n {b}_n{\cal K}_n^{\nu,\tau_n}(z_1,z_2)$ of the rescaled eigenvalue process ${Z}_n^{\tau_n,\nu}$ 
can be written
\begin{align}\label{kernelexplicit}
{\cal M}&_n^{\nu,\tau_n}({\zeta}_1,{\zeta}_2)\nonumber\\
=&\frac{\sqrt{|z_1||z_2|}n^{3/2}{a}_n {b}_n}{\pi^{5/2} \sqrt{(1-\tau_n^2)}}\left(\frac{\tau_n |z_1||z_2|}{z_1^2}\right)^{\nu}
\exp\left[n \left(2\log \tau_n + \frac{\overline{z}_2^2}{\tau_n}
-\frac{\left(|z_1^2|+|z_2^2|-\tau_n \operatorname{Re}(z_1^2+z_2^2)\right)}{(1-\tau_n^2)}\right)
\right]\nonumber \\
&\times
\oint_{\gamma_1}\ud w_1\oint_{\gamma_2}\ud w_2\left(\frac{w_2(w_1-1)}{(w_2-1)w_1}\right)^{\nu}
\frac{e^{n\left(f_n(w_2)-f_n(w_1) -\frac{w_2(\alpha_2-i\beta_2)}{\tau_n(w_2-1)}+\frac{w_1(\alpha_1+i\beta_1)}{\tau_n(w_1-1)}\right) }}
{(\tau_n^2w_2-w_1)(1-w_1)(1-w_2)}  (1+o(1)),
\end{align}
where $z_j:=c_n+{a}_n{\xi}_j+i {b}_n{\eta}_j$,
\begin{equation}\label{defalpha}
 \left\{ \begin{array}{ll} \alpha_j:=\operatorname{Re}(z_j^2)-c_n^2=2c_n{a}_n{\xi}_j+{a}_n^2{\xi}_j^2-{b}^2_n{\eta}_j^2 \\
\beta_j:=\operatorname{Im}(z_j^2)=2 b_n {\eta}_j(c_n+{a}_n{\xi_j}),
\end{array}\right.
\end{equation}
and where we have defined
\begin{equation}
f_n(w)=-\frac{ c_n^2}{\tau_n}\frac{w}{w-1}+\log w,
\end{equation}
with the branch cut of the logarithm chosen along the positive real axis, in order that $f_n$ be analytic in neighbourhoods of its saddle points.

The dominating contribution to the integral will come from saddle points of $f_n$,
given by $f'_n(w)=0$, or
\begin{equation*}
w^2+\left(\frac{c_n^2-2\tau_n}{\tau_n}\right)w+1=0,
\end{equation*}
which has two real solutions, $ w_-< -1<w_+<0 $.
\begin{remark}\label{scalealpha}
Note that, more generally, if $\nu$ scales with $n$, $\nu=\alpha n$, the integral can be evaluated in terms of saddle points of the function $g_n(w)=f_n(w)+\alpha(\log w-\log(w-1))$. In principle the analysis can be carried out for this case as well, however the edge of the spectrum $c_n$ then moves on a global scale and is only implicitly determined, which makes the calculations more difficult. However, the present analysis does apply for any fixed $\nu$.
\end{remark}

Explicitly, the two saddle points are given by
\begin{equation*}
w_{\pm}=1-\frac{c_n^2}{2\tau_n}\left(1 \mp \sqrt{1-\frac{4\tau_n}{c_n^2}}\right),
\end{equation*}
or, when expanded in $\delta_n/(1-\tau_n)^ 2 \ll 1$,
\begin{equation}
w_{-}=-\frac{1}{\tau_n}-\frac{\delta_n}{\tau_n(1-\tau_n^2)}\left(1-\tau_n^2(1+\tau_n)\frac{\delta_n}{(1-\tau_n)^2}
(1+o(1))\right)
\end{equation}
and \begin{equation}
w_{+}=-\tau_n+\frac{\delta_n\tau_n }{(1-\tau_n^2)}\left(1-(1+\tau_n)\frac{\delta_n}{(1-\tau_n)^2}(1+o(1))\right)
\end{equation}
respectively.
Since
\begin{equation}
f_n''(w)=-
\frac{2c_n^2}{\tau_n (w-1)^3}-\frac{1}{w^2}
,
\end{equation}
we find that
\begin{equation}\label{fbismin}
f_n''(w_-)=\frac{\tau_n^2 (1-\tau_n)}{(1+\tau_n)}
\left(1-\frac{2(1-\tau_n-\tau_n^2)}{(1+\tau_n)^2}\frac{\delta_n}{(1-\tau_n)^2}(1+o(1))
\right)>0
\end{equation}
and
\begin{equation}\label{fbisplus}
f_n''(w_+)=-\frac{(1-\tau_n)}{\tau_n^2(1+\tau_n)}
\left(1+\frac{2(1+\tau_n-\tau_n^2)}{(1+\tau_n)^2}\frac{\delta_n}{(1-\tau_n)^2}(1+o(1))
\right)<0,
\end{equation}
whereas
$ |f_n'''(w_{\pm})| \asymp 1$. 
Note that, here and in the following, correction terms depending on $\delta_n$ are identically zero unless $\sigma_n \to \infty$.

We will choose the contours so that $\gamma_1$ passes through $w_+$ and $\gamma_2$
through $w_-$ and are both tangent to vertical lines, giving the directions of steepest descent, by (\ref{fbismin}) and (\ref{fbisplus}).
Computing the asymptotics of the kernel, the main contribution will come from neighbourhoods of the saddle points, so neglecting small contributions away from the saddle points we can take the contours of integration as the lines  $w_++i\mathbb{R}$ and $w_-+i\mathbb{R}$ respectively and change variables $w_1-w_+=i{x} $,  $w_2-w_-=i{y}$. By expanding in powers of ${x}$ and ${y}$, we can find the asymptotics of the integral in (\ref{kernelexplicit}).
Before proceeding with the saddle point approximation,
we calculate asymptotics of the constant and subleading terms at the saddle points.

We note immediately that, provided $\tau_n$ is bounded away from $0$, the factor
\be \nn
\left(\frac{w_2(w_1-1)}{(w_2-1)w_1}\right)^{\nu}
\ee
in the integrand is approximately constant, taking the value
\begin{equation*}
\left(\frac{w_-(w_+-1)}{(w_--1)w_+}\right)^{\nu}=\tau_n^{-\nu}(1+o(1))
\end{equation*}  which cancels the pre-factor
\be \nn
\left(\frac{\tau_n |z_1||z_2|}{z_1^2}\right)^{\nu}=\tau_n^{\nu}(1+o(1)),
\ee
so that the limiting kernel will be independent of $\nu$.
Furthermore,
\be \nn
(w_1-1)(w_2-1)=\frac{(1+\tau_n)^2}{\tau_n}(1+o(1)),
\ee
and the remaining factor in the denominator becomes
\be \nn
\tau_n^2 w_2-w_1=\tau_n^2{x}-{y}-\frac{2\tau_n \delta_n}{(1-\tau_n^2)}(1+o(1)).
\ee
The leading order term of the exponent is
\begin{equation*} 
f_n(w_-)-f_n(w_+)=-2\log\tau_n-2\pi i -\frac{ c_n^2(1-\tau_n)}{\tau_n(1+\tau_n)}-\frac{\delta_n^2}{(1+\tau_n)^3(1-\tau_n)}(1+o(1)).
\end{equation*}
We will also need expansions of the sub-leading terms,
\begin{equation*}
\frac{w}{(w-1)}=
\frac{w_-}{(w_--1)}
-\frac{1}{(w_--1)^2}
i{x} +\mathcal{O}({x}^2),
\end{equation*}
where
\begin{equation*} 
\frac{w_-}{(w_--1)}
=\frac{1}{1+\tau_n}+\frac{\tau_n\delta_n}{(1+\tau_n)^3(1-\tau_n)}(1+o(1)),
\end{equation*}
and similarly
\begin{equation*} 
\frac{w_+}{(w_+-1)}
=\frac{\tau_n}{1+\tau_n}
-\frac{\tau_n\delta_n}{(1+\tau_n)^3(1-\tau_n)}(1+o(1)).
\end{equation*}

Collecting all the terms independent of the variables of integration from these calculations, and the exponential pre-factor, gives
\begin{align}\label{leadingterm}
2\log \tau_n& +\frac{\overline{z}_2^2}{\tau_n}
-\frac{|z_1|^2+|z_2|^2-\tau_n(\operatorname{Re}z_1^2+\operatorname{Re}z_2^2)}{(1-\tau_n^2)}
+f_n(w_-)-f_n(w_+)+\frac{w_+(\alpha_1+i\beta_1)}{\tau_n(w_+-1)}-\frac{w_-(\alpha_2-i\beta_2)}{\tau_n(w_--1)}
\nn\\
=&-\frac{\delta_n(\alpha_1+\alpha_2)}{(1+\tau_n)^3(1-\tau_n)}(1+o(1))
-\frac{(\beta_1^2+\beta_2^2)}{2c_n^2(1-\tau_n^2)}(1+\mathcal{O}(a_n))-\frac{\delta_n^2}{(1+\tau_n)^3(1-\tau_n)}(1+o(1))\nn\\
&+C_n(\beta_1-\beta_2)
-2\pi i +o(n^{-1}),
\end{align}
for some constant $C_n$.
The correlation kernel ${\cal M}_n^{\nu,\tau_n}(\zeta_1,\zeta_2)$ can
be replaced by any kernel of the form $e^{g_n(\zeta_1)-g_n(\zeta_2)}{\cal M}_n^{\nu,\tau_n}(\zeta_1,\zeta_2)$
without affecting the underlying point process since the determinants giving the correlation functions remain the same; we will thus 
consider the equivalent kernel
$e^{-C_n(\beta_1-\beta_2)}{\cal M}_n^{\nu,\tau_n}({\zeta}_1,{\zeta}_2)$,
which we still call ${\cal M}_n^{\nu,\tau_n}$ with a slight abuse of notation. Inserting (\ref{leadingterm}) into (\ref{kernelexplicit}) gives

\begin{align}\label{kernelexplicit1b}
&{\cal M}_n^{\nu,\tau_n}({\zeta}_1,{\zeta}_2)\nonumber\\
=&\frac{\tau_n n^{3/2}{a}_n {b}_ne^{-\frac{n\delta_n^2}{(1+\tau_n)^3(1-\tau_n)}}}{\pi^{5/2}(1+\tau_n)^{3/2}\sqrt{(1-\tau_n)}}
\exp\left[-n\left(\frac{\delta_n(\alpha_1+\alpha_2)}{(1+\tau_n)^3(1-\tau_n)}
+\frac{(\beta_1^2+\beta_2^2)}{2c_n^2(1-\tau_n^2)}\right)\right]\nonumber \\
&\times
\int_{\mathbb{R}}\ud {x}\int_{\mathbb{R}}\ud {y}
\frac{e^{n\left(f_n(w_-+i{y})-f_n(w_-)-f_n(w_++i{x})+f_n(w_+) +\frac{i{y}(\alpha_2-i\beta_2)}{\tau_n(w_--1)^2}-\frac{i{x}(\alpha_1+i\beta_1)}{\tau_n(w_+-1)^2}\right) }}
{i({x}-\tau_n^2{y})+\frac{2\tau_n \delta_n}{(1-\tau_n^2)}}  (1+o(1)),
\end{align}
where
\begin{equation}\label{expansion1b}
\frac{1}{(w_--1)^2}
=\frac{\tau_n^2}{(1+\tau_n)^2}\left(1-\frac{2\delta_n}{(1+\tau_n)^2(1-\tau_n)}(1+o(1))\right),
\end{equation}
and
\begin{equation}\label{expansion2b}
\frac{1}{(w_+-1)^2}
=\frac{1}{(1+\tau_n)^2}\left(1+\frac{2\tau_n\delta_n}{(1+\tau_n)^2(1-\tau_n)}(1+o(1))\right).
\end{equation}

\subsection{Case $\sigma_n \to \sigma<\infty $}
In this case we may choose $\delta_n=0$.
This means that, in order to get a convergent integral in (\ref{kernelexplicit1b}), we have to shift the contours of integration slightly into the complex plane. For the details of these considerations, we refer to the very similar arguments in \cite{Bender}; here we will formally integrate over the real line to clarify the presentation.
 Since $\tau_n$ tends to $1$ in this regime,
\begin{equation*}
f_n'''(w_{\pm})=f_n'''(-1)(1+o(1))=-\frac{1}{2}(1+o(1)).
\end{equation*}
After a change or variables,  $y=2(2n)^{-1/3}t$, $x=2(2n)^{-1/3}s$, in the integral in (\ref{kernelexplicit1b}), (\ref{fbismin}) and (\ref{fbisplus}) give
\be \nn
n(f_n(w_- + i y)-f_n(w_-))=-\frac{1}{2}\sigma_n^2 t^2+\frac{i}{3}t^3+\mathcal{O}(n^{-1/3}t^4),
\ee
and
\be \nn
n(f_n(w_+ + i x)-f_n(w_+))=-\frac{1}{2}\sigma_n^2 s^2-\frac{i}{3}s^3+\mathcal{O}(n^ {-1/3}s^4).
\ee
In the remaining expressions, we may directly insert $\tau_n=1$ and
(\ref{kernelexplicit1b}) becomes
\begin{align*}
{\cal M}_n^{\nu,\tau_n}({\zeta}_1,{\zeta}_2)=&\frac{{a}_n {b}_n n^{4/3}}{2^{2/3}\pi^{5/2}\sigma_n}
\exp\left[-\frac{n}{16(1-\tau_n)}(\beta_1^2+\beta_2^2)\right]\nonumber \\
&\times \int_{\mathbb{R}}\ud s\int_{\mathbb{R}}\ud t \frac{e^{-\frac{1}{2}\sigma_n^2(t^2+s^2)+\frac{i}{3}(t^3- s^3) +i 2^{-4/3}n^{2/3}\left((\alpha_2-i\beta_2)t -(\alpha_1+i\beta_1)s \right)}}
{i(s-t)}(1+o(1)).
\end{align*}
Recalling the definition (\ref{defalpha})  of our auxiliary variables $\alpha_j$ and $\beta_j$,
and choosing ${a}_n=(2n)^{-2/3}$ and ${b}_n= \sigma_n(2n)^{-2/3}$ gives
\begin{align*}
{\cal M}_n^{\nu,\tau_n}({\zeta}_1,{\zeta}_2)
=\frac{1}{4\pi^{5/2}}\int_{\mathbb{R}}\ud s\int_{\mathbb{R}}\ud t\frac{e^{ 
  -\frac{1}{2}(\sigma_n t- {\eta}_2)^2 +\frac{i}{3}t^3+i{\xi}_2 t -\frac{1}{2}(\sigma_n s-{\eta}_1)^2-\frac{i}{3}s^3-i{\xi}_1 s }}
{i(s-t)}(1+o(1))
\to {\cal K}^A_{\sigma}(\zeta_1,\zeta_2)
\end{align*}
as claimed.

For the weak convergence of point processes and convergence of the last particle distribution to follow, it now suffices that there for every $\xi_0$ is a uniform bound $B_{\sigma,\xi_0}(\zeta)> |{\cal M}_n^{\nu,\tau_n}({\zeta},{\zeta})|$ whenever $\operatorname{Re}(\zeta)>\xi_0$, such that $\int_{\operatorname{Re}(\zeta)>\xi_0}\ud^2 \zeta B_{\sigma,\xi_0}(\zeta)< \infty$. It can be shown that there is such a bound of the form $B_{\sigma, \xi_0}(\zeta)=C_{\sigma,\xi_0}e^{-\frac{1}{2}(\xi+\eta^2)}$. The argument parallels that of \cite{Bender}, and the details are omitted here.

\subsection{Case $\sigma_n \to \infty$}
This case requires a subtle choice of scaling parameters, to be determined by the analysis.
Change variables, $t=\sqrt{n f_n''(w_-)}\ y$ and $s=\sqrt{-n f_n''(w_+)}\ x$.
The imaginary part of the denominator in (\ref{kernelexplicit1b}) becomes
\be \nn
x-\tau_n^2y=i\tau_n\sqrt{\frac{(1+\tau_n)}{n(1-\tau_n)}}(1+o(1))(s-t),
\ee
so using the asymptotics (\ref{expansion1b}) and (\ref{expansion2b}), and noting that
\be \nn
\frac{1}{2c_n^2(1-\tau_n^2)}=\frac{1}{2(1+\tau_n)^3(1-\tau_n)}\left(1-\delta_n\frac{1}{(1+\tau_n)^2}\right)(1+o(1)),
\ee
we can write (\ref{kernelexplicit1b}) as
\begin{align}\label{kernelexplicit2}
&{\cal M}_n^{\nu,\tau_n}({\zeta}_1,{\zeta}_2)\nn \\
=&\frac{n {a}_n {b}_ne^{-\frac{n\delta_n^2}{(1+\tau_n)^3(1-\tau_n)}}}{\pi^{5/2} (1-\tau_n^2)}
\exp\left[-n\left(\frac{\delta_n(\alpha_1+\alpha_2)}{(1+\tau_n)^3(1-\tau_n)}
+\frac{(\beta_1^2+\beta_2^2)}{2(1+\tau_n)^3(1-\tau_n)}\left(1-\delta_n\frac{1}{(1+\tau_n)^2}\right)\right)\right]\nonumber \\
&\times
\int_{\mathbb{R}}\ud s\int_{\mathbb{R}}\ud t
 \frac{e^{-\frac{1}{2}(t^2+s^2)
 +A_n s+B_n t
 }}
{\rho_n+i(s-t)}(1+o(1)),
\end{align}
where
\be \nn
A_n:=\frac{\sqrt{n}}{\tau_n\sqrt{-f_n''(w_+)}(w_+-1)^2}(\beta_1-i\alpha_1),
\ee
\be \nn
B_n:=\frac{\sqrt{n}}{\tau_n\sqrt{f_n''(w_-)}(w_--1)^2}(\beta_2+i\alpha_2),
\ee
and
\be \nn
\rho_n:=\frac{2\delta_n\sqrt{n}}{\sqrt{(1-\tau_n^2)}(1+\tau_n)}>0.
\ee
We note that $\rho_n\asymp \sqrt{\log\sigma_n}\gg 1$, by assumption on $\sigma_n$.
Expanding, we find that
\be \nn
\operatorname{Im}\left(A_n-B_n\right)=-(\alpha_1+\alpha_2)\frac{\sqrt{n}}{\sqrt{(1-\tau_n)}(1+\tau_n)^{3/2}}(1+o(1))
\ll \rho_n,
\ee
and
\be \nn
\operatorname{Re}\left(A_n-B_n\right)=(\beta_1-\beta_2)\frac{\sqrt{n}}{\sqrt{(1-\tau_n)}(1+\tau_n)^{3/2}}(1+o(1))
\asymp \sigma_n^{3/2}(\log\sigma_n) ^{3/4} \gg \rho_n,
\ee
for $\beta_1 \neq \beta_2$.
To calculate the integral in (\ref{kernelexplicit2}), we write the denominator as an integral
and then compute the Gaussian integrals in $s$ and $t$, giving
\begin{align}\label{gumbelintegral}
\int_{\mathbb{R}}\ud s\int_{\mathbb{R}}\ud t\frac{e^{-\frac{1}{2}(t^2+s^2)+A_ns +B_nt}}{\rho_n+i(s-t)}&= \int_{\mathbb{R}}\ud s\int_{\mathbb{R}}\ud t \int_0^{\infty}\ud u \ e^{-\frac{1}{2}(t^2+s^2)+A_ns +B_nt-u(\rho_n+i(s-t))}\nn\\
&=2\pi\int_0^{\infty}\ud u \ e^{\frac{1}{2}(A_n-iu)^2+\frac{1}{2}(B_n+iu)^2-u\rho_n}\nn\\
&=2\pi e^{\frac{1}{2}(A_n^2+B_n^2)}\int_0^{\infty}\ud u \ e^{-u^2-u(\rho_n+i(A_n-B_n))}\nn\\
&= \frac{2\pi e^{\frac{1}{2}(A_n^2+B_n^2)}}{(\rho_n+i(A_n-B_n))}\left( 1+\mathcal{O} \left(\frac{1}{\rho_n^{2}}\right)\right)
\end{align}
where the last integral is calculated using integration by parts and a suitable splitting of the integral.
Inserting (\ref{gumbelintegral}) into (\ref{kernelexplicit2}) leads to some cancellation, and we are left with
\begin{align}\label{kernelexplicit4}
{\cal M}_n^{\nu,\tau_n}({\zeta}_1,{\zeta}_2)
=&\frac{\sqrt{(1+\tau_n)}\sqrt{n} {a}_n {b}_n
e^{-\frac{n\delta_n^2}{(1+\tau_n)^3(1-\tau_n)}}}
{\pi^{3/2}\sqrt{(1-\tau_n)}\delta_n(1+\frac{i}{2\delta_n}(\beta_1-\beta_2))}\nn\\
& \times \exp\left[-n\left(\frac{\delta_n(\alpha_1+\alpha_2)}{(1+\tau_n)^3(1-\tau_n)}
+\frac{(1+\tau_n^2)\delta_n(\beta_1^2+\beta_2^2)}{2(1+\tau_n)^5(1-\tau_n)^3}\right)\right](1+o(1))
\end{align}
where we have conjugated away a factor of the form $e^{C_n(\alpha_2\beta_2-\alpha_1\beta_1)}$ coming from (\ref{gumbelintegral}).
In terms of the original variables, the exponential terms of (\ref{kernelexplicit4}) become
\begin{align}\label{kernelexplicit5}
n\left(\frac{\delta_n(\alpha_1+\alpha_2)}{(1+\tau_n)^3(1-\tau_n)}
+\frac{(1+\tau_n^2)\delta_n(\beta_1^2+\beta_2^2)}{2(1+\tau_n)^5(1-\tau_n)^3}\right)=\frac{(2n)^{4/3}\delta_n a_n}{(1+\tau_n)^2\sigma_n^2}(\xi_1+\xi_2)
+\frac{\delta_n(2n)^2 b_n^2}{2(1+\tau_n)}(\eta_1^2+\eta_2^2)+o(1).
\end{align}
Now we choose $a_n$ and $b_n$ so that the coefficients in the right hand side of (\ref{kernelexplicit5}) tend to finite limits, say $1/2$, giving
\begin{equation}\label{choicea}
{a}_n = \frac{(1+\tau_n)^2\sigma_n^2}{2^{7/3}n^{4/3}\delta_n},
\end{equation}
and
\begin{equation}\label{choiceb}
{b}_n= \frac{\sqrt{(1+\tau_n)}\sigma_n^{3}}{2n\sqrt{\delta_n}}.
\end{equation}
Inserting into (\ref{kernelexplicit4}), we now choose $\delta_n$  so that we get a finite limit for the pre-factor, say
\begin{equation}\label{deltalimit}
\frac{2^{1/6}\sqrt{(1+\tau_n)}n^{2/3} {a}_n {b}_n e^{-\frac{n\delta_n^2}{(1+\tau_n)^3(1-\tau_n)}}}{\pi^{3/2}\sigma_n\delta_n}
=\frac{(1+\tau_n)^{3}}{2^{19/6}\pi^{3/2}}\frac{\sigma_n^4}{n^{5/3}\delta_n^{5/2}}
\exp\left[-\frac{2^{1/3}}{(1+\tau_n)^3}\frac{n^{4/3}\delta_n^2}{\sigma_n^2}\right]\to \frac{1}{\sqrt{\pi}}.
\end{equation}
It is easy to check that the choice
\begin{equation}\label{deltachoice}
\delta_n=\frac{(1+\tau_n)^{3/2}}{2^{1/6}}
\frac{\sigma_n }{\sqrt{6 \log \sigma_n}} n^{-2/3}
\left(3\log \sigma_n-\frac{5}{4}\log( 6\log \sigma_n)
-\log\left(2^{1/4}{\pi}(1+\tau_n)^{3/4}\right)\right)
\end{equation}
satisfies (\ref{deltalimit}).
 Note that this leads to a finite, non-zero limit for the kernel only on the diagonal, $\eta_1=\eta_2$; if $\eta_1\neq \eta_2$, we have  $|1+i(\beta_1-\beta_2)/(2\delta_n)|\gg 1$, implying that ${\cal M}_n^{\nu,\tau_n}({\zeta}_1,{\zeta}_2) \to 0$.
By definition,
\begin{equation*}
c_n=(1+\tau_n)\sqrt{1+\frac{\delta_n}{(1+\tau_n)^2}},\\
\end{equation*}
but since $a_n\gg \delta_n^ 2$, we may expand in $\delta_n$, neglect the second order term and choose
\begin{equation*}c_n\equiv (1+\tau_n)\left(1+\frac{\delta_n}{2(1+\tau_n)^2}\right)
\end{equation*}
Substituting back the leading term of (\ref{deltachoice}) into (\ref{choicea}) and (\ref{choiceb}), we can thus choose $a_n$ and $b_n$ given by (\ref{gumbelreal})  and (\ref{gumbelimaginary}) in the statement of the theorem.
With these choices, we get that
\begin{equation*}
{\cal M}_n^{\nu,\tau_n}({\zeta}_1,{\zeta}_2)\to \frac{1}{\sqrt{\pi}}e^{-{\eta}_1^2-{\xi}_1}\delta_{{\zeta}_1,{\zeta}_2},
\end{equation*}
the kernel of a two-dimensional Poisson process with intensity ${\pi}^ {-1/2}e^{-{\eta}^2-{\xi}}$.

Again, it can be shown that there is a uniform integrable bound  $B_{ \xi_0}(\zeta)=C_{\xi_0}e^{-\frac{1}{2}(\xi+\eta^2)}>|{\cal M}_n^{\nu,\tau_n}({\zeta},{\zeta})|$ for $\operatorname{Re}(\zeta)>\xi_0$, proving the convergence of the point process and last particle distribution to that of the limiting process.

\sect{Real integral representation of the interpolating Airy kernel}\label{proofla3}
\subsection{ Proof of Proposition \ref{Airyint}}
The proof of the alternative representation  (\ref{Airyrep2}) is a simple
matter of verification. Starting from  (\ref{Airyrep1b}), 
\begin{align}
\label{rep1}
\hat{\cal K}_{\sigma}^A(\zeta_1,\zeta_2)
=&  \frac{e^{-\frac{1}{2\sigma^2}(\eta_1^2+\eta_2^2)}}{4\pi^{5/2}\sigma}\int_{\Gamma}\ud u\int_{\Gamma}\ud v \int_0^\infty\ud t \
e^{it(u+v)-\frac{1}{2}\sigma^2
    u^2+\frac{i}{3}u^3+i\zeta_1 u-\frac{1}{2}\sigma^2
    v^2+\frac{i}{3}v^3+i\overline{\zeta}_2 v},
  \end{align}
where we have written the denominator as a convergent integral.
Putting
$
z_j:= \zeta_j+\sigma^4/4
$,
the exponent in  (\ref{rep1}) can be trivially rewritten
as
\begin{align}\label{exponent}
it(u+v)-&\frac{1}{2}\sigma^2
    u^2+\frac{i}{3}u^3+i\zeta_1 u-\frac{1}{2}\sigma^2
    v^2+\frac{i}{3}v^3+i\overline{\zeta}_2 v \nn\\
=&it(u+v) -\frac{\sigma^6}{12}
+\frac{\sigma^2}{2}\left({z}_1+\overline{z}_2
\right)
\nn \\
&+i{z}_1\left(u+i\frac{\sigma^2}{2}\right)
+\frac{i}{3}\left(u+i\frac{\sigma^2}{2}\right)^3 + i\overline{z}_2
\left(v+i\frac{\sigma^2}{2}\right)
+\frac{i}{3}\left(v+i\frac{\sigma^2}{2}\right)^3
.
\nn
\end{align}
Using the well known integral representation of the Airy function,
\be \nn
\Ai(x)= \frac{1}{2\pi}\int_{\Gamma}\ud u \ e^ {ixu+\frac{i}{3}u^3}
\ee
we obtain (\ref{Airyrep2}), upon substituting $q=u+i\frac{\sigma^2}{2}$ and
$p=v+i\frac{\sigma^2}{2}$. 
\subsection{Decay of density}
The asymptotic eigenvalue density when $\sigma_n\to \infty$ can be computed directly as the large $n$ limit of the density, as expressed in (\ref{errorfunctiondecay}), by an appropriate steepest descent analysis of ${\cal K}_n^{\nu,\tau_n}$. We will instead give a simple heuristic argument by calculating the large $\sigma$ limit of the density of the interpolating Airy process, after rescaling  by a factor $\sigma$, and recover the same expected decay of the density. Roughly speaking this amounts to saying that the same limit is obtained on the one hand if  $\sigma_n\to \infty$ with $n$, and on the other if the large $n$ limit is taken with $\sigma$ remaining constant, and then letting $\sigma$ tend to infinity.

Changing variables $t=\sigma v$ in the integral in (\ref{Airyrep2}), we obtain
\begin{equation*}
\hat \rho_\sigma(\zeta):=\sigma^2\hat{\cal K}_{\sigma}^A(\sigma \zeta,\sigma \zeta)
=\frac{\sigma^2 e^{\frac16\sigma^6+\sigma^3\xi-\eta^2
}}{\sqrt{\pi}}
\int_{0}^\infty \ud v \ e^{v\sigma^3}\
{\rm Ai}(R e^{i\theta})
{\rm Ai}(R e^{-i\theta}),
\end{equation*}
where $R=\sigma \sqrt{(\sigma^3/4+v+\xi)^2 +\eta^2}$ and $\tan \theta=\eta(\sigma^3/4+v+\xi)^{-1}$. Using the large argument asymptotics of the Airy function, (\ref{Airyasymptotics}), this can be written
\begin{equation}\label{erfcint}
\hat \rho_\sigma(\zeta)=
\frac{e^{\frac16\sigma^6+\sigma^3\xi-\eta^2} }
{2  \pi^{3/2}}
\int_{0}^\infty \ud v \
e^{ v\sigma^3-\frac{4}{3}R^{3/2}\cos\left(\frac{3\theta}{2}\right)}(1+o(1)).
\end{equation}
Since $\theta \ll 1$, we can expand in $\theta$, and replace the highest order term with its constant value $\theta_0=4\eta/\sigma^3$, leading to the approximate integrand $e^{g_\sigma(v)}$, where
\begin{equation*}
g_{\sigma}(v)=v\sigma^3-\frac{\sigma^{6}}{6}\left(1+\frac{4(v+\xi)}{\sigma^3}\right)^{3/2}\left(1-\frac{6\eta^2}{\sigma^6}\right),
\end{equation*}
defined for $v>-\sigma^3/4-\xi$.
For fixed $\zeta$, $g_{\sigma}$ attains its maximum at
$v_0\approx -\xi$, which stays of order $1$ as $\sigma \to \infty$. Since $g_{\sigma}''(v)$ is also of order $1$ for $v\ll \sigma$, and $g_{\sigma}'''(v)=\mathcal{O}(\sigma^{-3})$,  expanding around $v=0$ up to second order in $v$ will give
the main contribution to the integral in (\ref{erfcint}).
Neglecting terms of order $\sigma^{-1}$, we calculate
$g_{\sigma}(0)
=-\sigma^6/6-\sigma^3\xi+\eta^2-\xi^2$, $g_{\sigma}'(0)=-2\xi$, and
$g_{\sigma}''(0)=-2$,
 which gives
\begin{equation*}
\hat \rho_\sigma(\zeta)\to \frac{e^{\frac16\sigma^6+\sigma^3\xi-\eta^2}}
{2  \pi^{3/2}}
\int_{0}^{\infty} \ud v \
e^{-\frac{\sigma^6}{6}-\sigma^3\xi+\eta^2-\xi^2-2\xi v-v^2 }= \frac{1}
{2  \pi^{3/2}}
\int_{0}^{\infty} \ud v \
e^{-(\xi+v)^2 }= \frac{1}{4\pi} \operatorname{erfc}(\xi)
\end{equation*}
as $\sigma \to \infty$.

\sect{Open problems}

There are several open problems left for future work.
First of all, regarding universality, it would be very interesting to
show that the interpolating kernel we have found also appears in the scaling limit
for models with more general non-Gaussian weight functions,
for instance the class of harmonic potentials.

A second question is regarding the other symmetry classes $\beta=1,4$,
where we expect different interpolating Airy kernels, generalizing the
corresponding Airy kernels for real eigenvalues. Perhaps the relations
between the corresponding kernels pointed out recently in \cite{APSo}
could open a simple way to obtain these.

A somewhat more ambitious task is to investigate whether or not the
relation between the Tracy-Widom distribution and the Painlev\'e II
solution has an analogue in the complex plane. It is possible that this
will require a better understanding of the corresponding
Riemann-Hilbert problem mentioned earlier.\\[2ex]

Acknowledgments:
This work has been supported partly by
European Network ENRAGE MRTN-CT-2004-005616 (G.A.).
The second author was supported by K.U. Leuven research grant OT/08/33, and
the Belgian Interuniversity Attraction Pole P06/02. We would like to thank to organizers of the CRM
workshop ``Random matrices, related topics and applications''
in Montreal
where this collaboration was initiated.
We are also indebted to Boris Khoruzhenko for useful discussions and
sharing the manuscript \cite{KS09} prior to publication.


\end{document}